\pdfoutput=1

\documentclass[10pt,conference]{IEEEtran}
\IEEEoverridecommandlockouts

\usepackage[utf8]{inputenc}
\usepackage[T1]{fontenc}
\usepackage{cite}
\usepackage{amsmath,amssymb,amsfonts}
\usepackage{graphicx}
\usepackage[dvipsnames,rgb,table]{xcolor}
\usepackage{listings}
\usepackage{microtype}
\usepackage{url}
\usepackage[scaled=0.8]{beramono}

\definecolor{navy-blue}{cmyk}{1.0,1.0,0.0,0.5}
\definecolor{grey}{rgb}{0.6,0.6,0.6}
\definecolor{eng}{cmyk}{0.14,0.71,1,0.03}
\definecolor{um}{rgb}{0.686,0.126,0.106}
\definecolor{mygreen}{rgb}{0,0.6,0}
\definecolor{mymauve}{rgb}{0.58,0,0.82}
\xdefinecolor{clrRE}{HTML}{0F0057}

\usepackage[
  pdftex,
  colorlinks = true,
  bookmarks = false,
  linkcolor = clrRE,
  citecolor = clrRE,
  urlcolor = clrRE
]{hyperref}

\lstdefinelanguage{hpl}{
    keywords = {globally, after, until, within, some, no, causes, requires, forbids, to, not, in, and, or, forall, as, ms, True, False},
    moredelim={[is][\color{eng}]{@@}{@@}}
}

\newcommand{\rosname}[1]{{\color{eng}\texttt{#1}}}
\newcommand{\langkw}[1]{{\color{navy-blue}\bfseries\texttt{#1}}}

\newcommand{\inlinepy}[1]{\lstinline|#1|}
\newcommand{\inlinehpl}[1]{\lstinline[language=hpl]!#1!}

\def\CC{{C\nolinebreak[4]\hspace{-.05em}\raisebox{.35ex}{\scriptsize\textbf ++}}}

\begin{document}


\title{The High-Assurance ROS Framework
\thanks{This work is financed by the ERDF - European Regional Development Fund through the Operational Programme for Competitiveness and Internationalisation - COMPETE 2020 Programme and by National Funds through the Portuguese funding agency, FCT - Fundação para a Ciência e a Tecnologia within project PTDC/CCI-INF/29583/2017 (POCI-01-0145-FEDER-029583).}
}

\author{\IEEEauthorblockN{André Santos}
\IEEEauthorblockA{\textit{High-Assurance Software Laboratory} \\
\textit{INESC TEC \& University of Minho}\\
Braga, Portugal \\
0000-0002-1985-8264}
\and
\IEEEauthorblockN{Alcino Cunha}
\IEEEauthorblockA{\textit{High-Assurance Software Laboratory} \\
\textit{INESC TEC \& University of Minho}\\
Braga, Portugal \\
0000-0002-2714-8027}
\and
\IEEEauthorblockN{Nuno Macedo}
\IEEEauthorblockA{\textit{High-Assurance Software Laboratory} \\
\textit{INESC TEC \& University of Porto}\\
Porto, Portugal \\
0000-0002-4817-948X}
}

\maketitle


\begin{abstract}
This tool paper presents the High-Assurance ROS (HAROS) framework.
HAROS is a framework for the analysis and quality improvement of robotics software developed using the popular Robot Operating System (ROS).
It builds on a static analysis foundation to automatically extract models from the source code.
Such models are later used to enable other sorts of analyses, such as Model Checking, Runtime Verification, and Property-based Testing.
It has been applied to multiple real-world examples, helping developers find and correct various issues.\footnote{%
\textcopyright{}2021 IEEE. %
Personal use of this material is permitted. %
Permission from IEEE must be obtained for all other uses, in any current or future media, including reprinting/republishing this material for advertising or promotional purposes, creating new collective works, for resale or redistribution to servers or lists, or reuse of any copyrighted component of this work in other works.\\
Postprint, after review.\\
To appear in: Proceedings of the 3rd International Workshop on Robotics Software Engineering (RoSE@ICSE 2021), Madrid, Spain, 4 pages
}
\end{abstract}


\section{Introduction}

There is little doubt that today's robots are capable of incredible feats.
Innovation is constant, expectations are high, and the responsibilities we place on robots are ever increasing.
Robots are the new definition of safety-critical devices.
But, what can we say about their overall software quality?

Primitive robot systems, much like any relatively new technology, were built mostly in an ad hoc fashion.
Over time, some middlewares thrived and became standards among practitioners.
Among them, the Robot Operating System~\cite{QuigleyCGFFLWN:09} (ROS) became an established backbone of open-source robotic software development~\cite{AlamiDW:18, GarciaDLSKB:19}.
Initiatives such as the ROSIN EU Horizon 2020 project\footnote{\url{https://www.rosin-project.eu/}} and the ROS Quality Assurance Working Group\footnote{\url{https://discourse.ros.org/c/quality/}} are fundamental steps in promoting established software engineering practices, such as Model-driven Engineering.
Despite their efforts, adoption by the general ROS community is still a slow work in progress.
Traditional, code-first development is the norm; software models are nowhere to be seen.

When confronted with the question \emph{``Does my robot do what it is supposed to do, reliably?''}, we want to be able to answer it with some degree of certainty.
It is well known in the software engineering field that this is not an easy question to answer.
Ultimately, a system should only be deemed safe, or dependable, if a set of critical properties is considered satisfied.
These properties can be checked with a number of techniques, such as Formal Verification, Model Checking, Runtime Verification, and more, but using one technique in isolation might not be sufficient.
Moreover, it is often the case that these techniques are only able to address relatively low-level properties of small units of software, not the system as a whole.
Ideally, we want to specify high-level, system-wide dependability properties and have a direct means of showing that they hold.
This is where dependability cases~\cite{Jackson:09} come in.

A dependability case is an end-to-end argument, supported by concrete evidence, that a system satisfies a given property.
The argument spans the system both horizontally, considering various inputs and outputs, and vertically, from design level down to the source code.
Properties can be broken down and evidence for each sub-property can be harnessed using different verification techniques.
It is, thus, a way to systematically combine verification techniques and play to the strengths of each, as needed.
Still, using the individual verification techniques requires expertise that one rarely finds in the common ROS developer~\cite{AlamiDW:18}.
This is the problem we address.

\smallskip
\emph{%
    How can existing software analysis techniques and tools be used by non-experts, to improve the quality of ROS applications and to provide the basis for dependability cases?%
}
\smallskip

Standard software analysis techniques can be employed behind an interface that caters to ROS roboticists.
Namely, an interface that \emph{(i)} takes source code as input, \emph{(ii)} reverse engineers formal models as needed, and that \emph{(iii)} uses a high-level property specification language that addresses ROS concepts directly.
These are the guiding principles behind the High-Assurance ROS framework (HAROS)~\cite{SantosCML:16}, found at:\\
\centerline{\texttt{\url{https://github.com/git-afsantos/haros}}}

In this paper we explain the analysis workflow of HAROS (Section~\ref{sec:haros}) and show how it combines multiple verification techniques (Section~\ref{sec:plugins}).
We compare it to other relevant tools in Section~\ref{sec:related}.
Lastly, in Section~\ref{sec:conclusion} we summarize our work, some experimental results with real-world case studies and some directions for future work.
Compared with earlier pub\-li\-ca\-tions~\cite{SantosCML:16, SantosCM:19}, this paper provides more up-to-date figures, code fragments and an up-to-date report about the current state and future plans for the project.

\section{Tool Overview and Workflow} \label{sec:haros}

HAROS is specifically designed for the analysis of ROS software.
One of its core features is a metamodel describing how ROS software is structured, both at runtime and in a file system~\cite{SantosCM:19}.
At runtime, a ROS system is composed of a network of processes, called \emph{nodes}, communicating via message-passing.
Messages can be exchanged following a publisher-subscriber paradigm (\emph{ROS topics}) or a client-server paradigm (\emph{ROS services}).
There is also a shared key-value store where arbitrary data is read and written (\emph{ROS parameters}).
At the file system level, ROS software is distributed in units called \emph{packages}, which contain a variety of files, such as CMake build files, \CC\ and Python source code, system deployment scripts (\emph{launch files}) and message type definition files, to name a few.

In practice, HAROS is divided into two components: the \emph{analyser}, a Python console application that does the bulk of the work (sometimes simply called HAROS); and the \emph{visualizer} (or \emph{viz}, for short) that handles interactive reports using web technologies.
HAROS comes also with a companion repository\footnote{\url{https://github.com/git-afsantos/haros_tutorials}} containing a minimalistic ROS application.
It consists of a driver for a fictitious robot (Fictibot), a random walker controller and a multiplexer that sorts velocity commands by priority.
We detail both components of HAROS in the remainder of this section, using Fictibot as the running example.

\subsection{HAROS Analyser}

\begin{figure}
\centering
\includegraphics[width=\columnwidth]{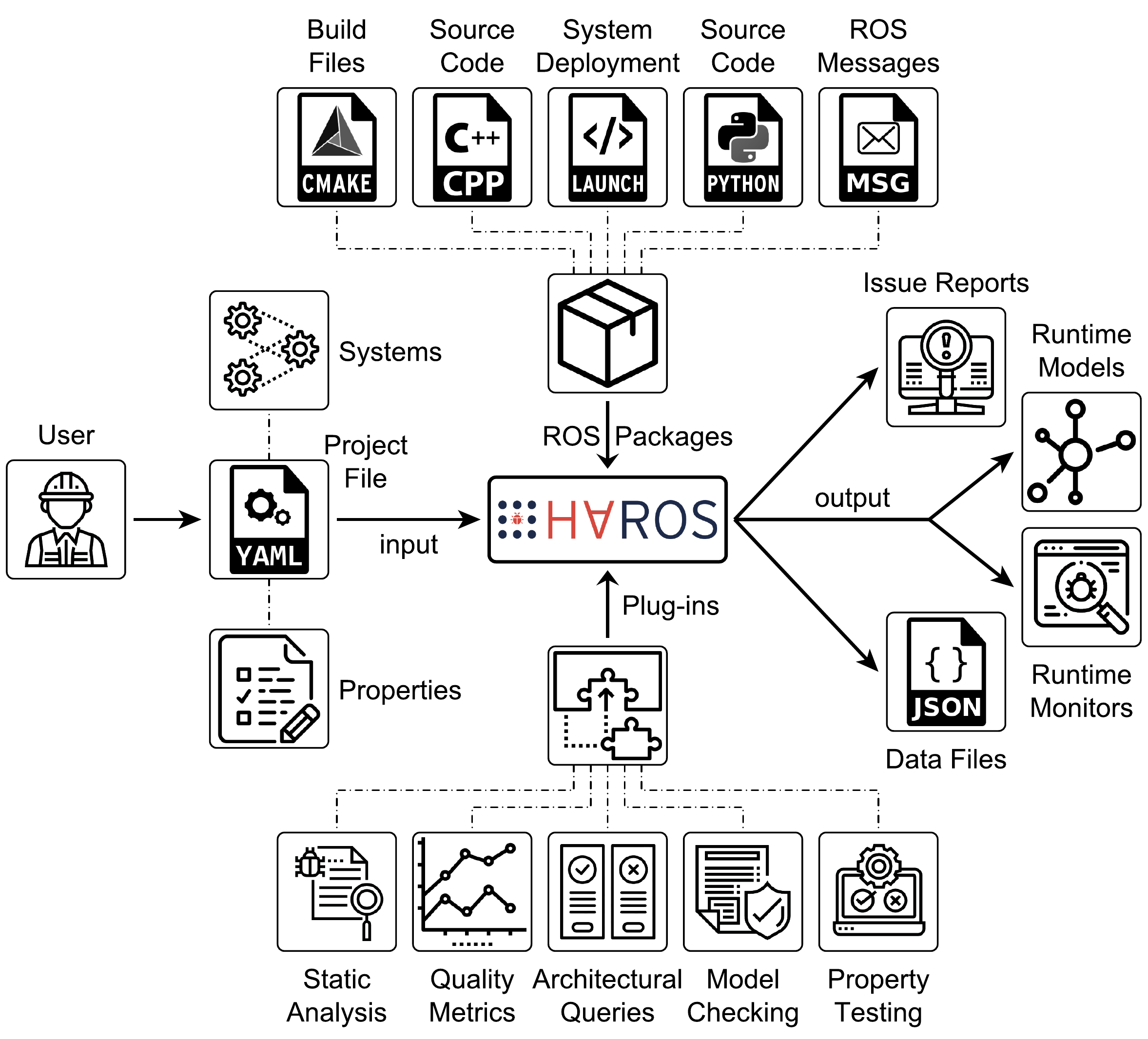}
\caption{Workflow of the HAROS static analysis framework.}
\label{fig:haros:haros}
\end{figure}

The typical workflow of the HAROS analyser is shown in Fig.~\ref{fig:haros:haros}.
It starts by processing a user-provided YAML \emph{project file}, such as the one shown in Fig.~\ref{fig:haros:hints}, whose primary purpose is to define analysis targets.
In this file, users specify \emph{configurations} -- lists of launch files that represent concrete robotic systems or applications.
This is necessary because ROS does not have a well-defined concept of system or application.
In this example, we can see that only 4 packages should be considered for analysis, and only one configuration is defined, \texttt{multiplex}, consisting of a single launch file.
Configurations convey only architectural information (nodes, topics, etc.).
They can be annotated with architectural and behavioural properties to be checked during the proper analysis stage (not shown).

HAROS provides a minimalistic, message-based specification language\footnote{\url{https://github.com/git-afsantos/hpl-specs}} for behavioural properties (HPL)~\cite{CarvalhoCMS:20}, although others can be plugged-in.
With HPL, users can specify safety properties, e.g., `\inlinehpl{globally: no @@/bumper@@ \{data < 0 or data > 7\}}', or liveness properties, e.g., `\inlinehpl{globally: @@/bumper@@ causes @@/stop_cmd@@}'.
The former restricts valid values for bumper states, while the latter states that a bumper message causes the system to respond with a stop command, eventually.

After processing command-line arguments and the project file, HAROS proceeds to an \emph{indexing} stage, during which it builds instances of its metamodel.
It locates packages and files in the file system, and then extracts information from these artefacts.
If enabled, it parses the launch files corresponding to each configuration at this point.
These dictate which nodes (binary executables) make up the target system, and how they are orchestrated.
The CMake files are used afterwards, to associate binaries with source code.
Lastly, \CC\ and Python files are parsed, to identify topics, services and parameters that nodes use at runtime.
This step by step procedure yields, in the end, a complete model, from source artefacts to the network of runtime entities, without ever executing code.

Naturally, given the complexity of \CC\ and Python code, the extraction process is not complete.
It covers most of the common use cases, but some elements of the extensive ROS API are not yet covered (e.g., abstractions provided by packages such as \texttt{tf2}).
Also, in some cases, it is impossible to determine dynamic values ahead of time.
HAROS addresses this by taking into account (optional) user-provided extraction hints specified in the project file (shown in Fig.~\ref{fig:haros:hints}).
Hints are partial, i.e., it is not necessary to specify the full system.
In the example, hints state that the node \rosname{/ficticontrol} should publish messages of type \texttt{std\_msgs/Float64} on topic \rosname{/controller\_cmd}.

\begin{figure}
\begin{lstlisting}[
    basicstyle=\ttfamily\footnotesize,
    moredelim={[is][\bfseries\color{navy-blue}]{@@}{@@}},
    abovecaptionskip=0pt,
    belowcaptionskip=0pt,
    belowskip=-1em
]
@@project@@: Fictibot
@@packages@@: ["fictibot_drivers", "fictibot_msgs", "fictibot_controller", "fictibot_multiplex"]
@@configurations@@:
  @@multiplex@@:
    @@launch@@: ["fictibot_controller/launch/multiplexer.launch"]
    @@hints@@:
      @@nodes@@:
        @@/ficticontrol@@:
          @@publishers@@:
            - @@topic@@: "/controller_cmd"
              @@msg_type@@: "std_msgs/Float64"
\end{lstlisting}
\caption{Project file with a configuration and extraction hints.}
\label{fig:haros:hints}
\end{figure}

Once all models are instantiated, the analysis step begins.
Despite its name, the HAROS analyser simply delegates analyses to any installed plug-ins.
This accomplishes three goals: \emph{(i)} reuse of existing tools, if possible (plug-ins can be simple wrappers for other tools); \emph{(ii)} adaptability to various use cases (not all users want all analysis capabilities); and \emph{(iii)} homogeneity of issue reports (all plug-ins register their results via a single interface).
Plug-ins are installed independently of the main tool, and can be blacklisted via user-provided arguments.
Section~\ref{sec:plugins} presents some of the available plug-ins.

Lastly, in the \emph{reporting} stage, HAROS exports JSON data files containing, e.g., extracted runtime models and analysis issues aggregated by package.
Plug-ins can also generate ar\-bi\-trary files at this stage (e.g., source code), for later use.

\subsection{HAROS Visualizer}

\begin{figure}
\centering
\includegraphics[width=0.95\columnwidth]{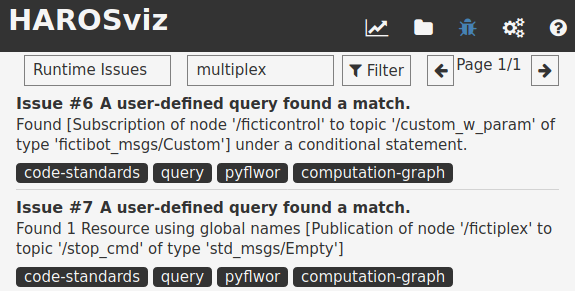}
\caption{Issue listing with the HAROS visualizer.}
\label{fig:haros:viz1}
\end{figure}

\begin{figure}
\centering
\includegraphics[width=0.95\columnwidth]{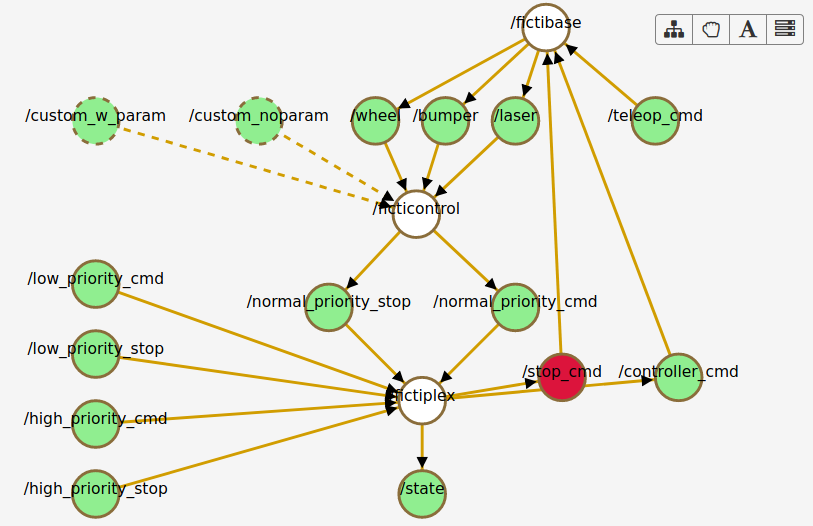}
\caption{Runtime model with issue highlights in the HAROS visualizer.}
\label{fig:haros:viz2}
\end{figure}

The HAROS visualizer produces an interactive report based on the exported JSON data files.
It defaults to a dashboard page where summary data is provided for a selected project.
This page sorts the information in three panels: source code statistics (e.g., number of packages and files), analysis statistics (e.g., total number of issues) and history of several metrics.
Other pages include: a package overview, where packages are drawn in a dependency graph; a list of issues reported by plug-ins, organized by category (Fig.~\ref{fig:haros:viz1} shows issues of the \texttt{multiplex} configuration); and interactive models of the extracted runtime configurations (Fig.~\ref{fig:haros:viz2} shows the \texttt{multiplex} configuration).
Conditional entities (e.g., under \langkw{if}) in the model are drawn in dashed lines (top left in the diagram).
Subjects of issues can also be highlighted (bottom right, shown in red).

\section{HAROS Analysis Plug-ins} \label{sec:plugins}

HAROS users are able to create plug-ins to fit their needs.
Here we list some plug-ins that we know to be freely available.

\subsection{Static Analysis}

These plug-ins\footnote{\url{https://github.com/git-afsantos/haros_plugins}} take in \CC\ and Python code directly.
They are based on existing tools for general-purpose analyses.
Some tools, often called \emph{linters} (e.g., \texttt{cpplint} and \texttt{pylint}), detect various issues, mostly related to formatting, according to some coding standards.
Others, like \texttt{cppcheck}, detect small and common bugs, such as uninitialized variables and out-of-bounds errors.
Yet others, like \texttt{lizard} and \texttt{radon}, measure a number of quality metrics, such as cyclomatic complexity, and report violations of these metrics against popular quality standards.

\subsection{Architectural Queries}

This plug-in\footnote{\url{https://github.com/git-afsantos/haros-plugin-pyflwor}} is a query engine over the extracted runtime models that checks user-defined structural rules.
Basic use cases include ensuring that every topic has at most one publisher (a common guideline in many systems), or detecting the use of conditional publishers and subscribers (Issue \#6, shown in Fig.~\ref{fig:haros:viz1}).
The latter is given by the pattern `\inlinepy{@!nodes/publishers!@[self.conditions] \| @!nodes/subscribers!@[self.conditions]}', that matches publishers or subscribers with associated conditions (e.g., an \langkw{if}).
A more complex example is a compile-time type-checking system for ROS topics and services, a feature that ROS lacks.

\subsection{Model Checking}

The main goal of this plug-in\footnote{\url{https://github.com/nmacedo/haros_plugin_mc}} is to verify system-wide behavioural properties in ROS applications, i.e., properties that span a whole configuration, rather than single node behaviour.
Node behaviour must be axiomatized with additional properties, to make system-wide verification possible (e.g., what the node publishes, dependencies between messages).
Verification of node-specific behaviour is delegated to other plug-ins, such as the testing plug-in we present next.
The proposed technique formalizes models and HPL properties in Electrum\footnote{\url{http://haslab.github.io/Electrum/}}~\cite{BrunelCCM:18}, a model checker for relational first-order temporal specifications.

\subsection{Testing and Runtime Verification}

Taking in HPL behavioural specifications, this plug-in\footnote{\url{https://github.com/git-afsantos/haros-plugin-pbt-gen}} generates runtime monitors and property-based tests.
It converts each property into a testing \emph{schema} -- a strategy to narrow down traces of messages that falsify the input property.
Then, it uses Hypothesis\footnote{\url{https://hypothesis.works/}} to convert schemas into input generators, and to explore the input space repeatedly.
Counterexample message traces are detected with runtime monitors, minimized (or \emph{shrunk}) with Hypothesis, and then presented to the user.

\section{Related Work} \label{sec:related}

To the best of our knowledge, there are no existing tools directly comparable to HAROS, in terms of performing ROS-specific property checks using a variety of analysis techniques.
There are, however, many approaches focusing on a subset of the problems HAROS addresses.
For instance, Statick\footnote{\url{https://github.com/sscpac/statick}} is similar to HAROS, in concept.
It is a plug-in based tool with domain knowledge about ROS packages.
It integrates a number of analysis tools and unifies their reports, but it focuses only on ROS-agnostic static analyses (e.g., linters and quality metrics).

Automatic model extraction from ROS source code, using static analysis has been done in \cite{PurandareDED:12} and \cite{SharmaED:17}.
However, these approaches focus only on the publisher-subscriber aspect, neglecting ROS services and parameters.
They are also not aimed at enabling other general-purpose analyses; HAROS not only builds models but makes them available to users.

Witte and Tichy~\cite{WitteT:18} propose a process to extract runtime models that uses static analysis for launch files and dynamic analysis for node interfaces.
Nodes run within a sandboxed environment that intercepts calls to build topics and services.
A limitation of this approach, in addition to executing code, is the assumption that nodes follow a standard life cycle, in which topics and services are created during set up.
This is not necessarily true; resources can be created at any time.

Verification of behavioural properties in ROS is commonly tackled via runtime verification.
Some of the most prominent tools are ROSRV~\cite{HuangEZMLSR:14}, ROSMonitoring~\cite{FerrandoC0AFM:20} and DeRoS~\cite{AdamLJS:16}.
The first does not offer a property specification language; properties are programmed manually.
The second uses a relatively low-level, domain-agnostic specification language based on regular expressions and events as JSON objects.
The third provides a language to specify both architecture and properties; monitors are capable of enforcing temporal properties and safety actions, albeit without formal semantics.

\section{Conclusion} \label{sec:conclusion}

This paper presented HAROS, a plug-in driven framework to analyse properties of ROS systems.
One of its core features is the semi-automatic extraction of runtime architectures at compile time.
The extracted models not only offer visual feedback to developers, but also enable model-based analyses via plug-ins, such as verifying structural properties via queries, verifying behavioural properties via model checking, or using the models to generate property-based tests.

HAROS has been tested on multiple real-world case studies, including academic examples~\cite{SantosCM:19} (e.g., TurtleBot2\footnote{\url{https://www.turtlebot.com/turtlebot2/}}), commercial products~\cite{GarciaDLSKB:19} (e.g., Care-O-bot~4\footnote{\url{https://www.care-o-bot.de/en/care-o-bot-4.html}}) and industrial robots~\cite{CarvalhoCMS:20, NetoASSV:19}.
Despite its limitations, especially regarding plug-ins based on behavioural properties, we have observed good results, overall.
For instance, the model extractor has required hints only for one in every ten entities, on average.
In addition, our property-based testing plug-in has been able to unveil safety bugs in a hillside vineyard agricultural robot, one of the industrial case studies, that have been since reported and fixed.
A repository of HAROS case studies and tangible artefacts that are possible with the presented workflow can be found at:\\
\centerline{\texttt{\url{https://github.com/git-afsantos/haros-case-studies}}}

Regarding the future of HAROS, firstly, we intend to alleviate extraction hints, extending HAROS's domain knowledge to cover standard ROS packages and the new, fast-developing ROS2.
Then, we intend to reduce the number of user-specified configurations, integrating Feature Models and variability-aware analyses.
Lastly, we plan on developing new plug-ins, e.g., to verify HPL intra-node properties based on control flow analysis and software model checking, or to profile energy consumption.

\section*{Data Availability}

HAROS, the HPL specification language and the repository of case studies and analysis artefacts are openly available in \texttt{zenodo.org} at the following addresses, respectively:\\
\centerline{\texttt{\url{https://doi.org/10.5281/zenodo.4569751}}}\\
\centerline{\texttt{\url{https://doi.org/10.5281/zenodo.4570107}}}\\
\centerline{\texttt{\url{https://doi.org/10.5281/zenodo.4569749}}}

\bibliographystyle{IEEEtran}
\bibliography{ms}

\begin{thebibliography}{10}
\providecommand{\url}[1]{#1}
\csname url@samestyle\endcsname
\providecommand{\newblock}{\relax}
\providecommand{\bibinfo}[2]{#2}
\providecommand{\BIBentrySTDinterwordspacing}{\spaceskip=0pt\relax}
\providecommand{\BIBentryALTinterwordstretchfactor}{4}
\providecommand{\BIBentryALTinterwordspacing}{\spaceskip=\fontdimen2\font plus
\BIBentryALTinterwordstretchfactor\fontdimen3\font minus
  \fontdimen4\font\relax}
\providecommand{\BIBforeignlanguage}[2]{{%
\expandafter\ifx\csname l@#1\endcsname\relax
\typeout{** WARNING: IEEEtran.bst: No hyphenation pattern has been}%
\typeout{** loaded for the language `#1'. Using the pattern for}%
\typeout{** the default language instead.}%
\else
\language=\csname l@#1\endcsname
\fi
#2}}
\providecommand{\BIBdecl}{\relax}
\BIBdecl

\bibitem{QuigleyCGFFLWN:09}
M.~Quigley, K.~Conley, B.~P. Gerkey, J.~Faust, T.~Foote, J.~Leibs, R.~Wheeler,
  and A.~Y. Ng, ``{ROS}: An open-source {Robot Operating System},'' in
  \emph{{ICRA} Workshop on Open Source Software}, 2009.

\bibitem{AlamiDW:18}
A.~Alami, Y.~Dittrich, and A.~Wasowski, ``Influencers of quality assurance in
  an open source community,'' in \emph{Int. Workshop on Cooperative and Human
  Aspects of Software Engineering ({ICSE})}, 2018, pp. 61--68.

\bibitem{GarciaDLSKB:19}
N.~H. Garcia, L.~Delval, M.~L{\"{u}}dtke, A.~Santos, B.~Kahl, and M.~Bordignon,
  ``Bootstrapping {MDE} development from {ROS} manual code - part 2: Model
  generation,'' in \emph{{ACM/IEEE} Int. Conf. on Model Driven Engineering
  Languages and Systems ({MODELS})}.\hskip 1em plus 0.5em minus 0.4em\relax
  {IEEE}, 2019, pp. 95--105.

\bibitem{Jackson:09}
D.~Jackson, ``A direct path to dependable software,'' \emph{Communications of
  the {ACM}}, vol.~52, no.~4, pp. 78--88, 2009.

\bibitem{SantosCML:16}
A.~Santos, A.~Cunha, N.~Macedo, and C.~Louren{\c{c}}o, ``A framework for
  quality assessment of {ROS} repositories,'' in \emph{{IEEE/RSJ} Int. Conf. on
  Intelligent Robots and Systems ({IROS})}, 2016, pp. 4491--4496.

\bibitem{SantosCM:19}
A.~Santos, A.~Cunha, and N.~Macedo, ``Static-time extraction and analysis of
  the {ROS} computation graph,'' in \emph{{IEEE} Int. Conf. on Robotic
  Computing ({IRC})}, 2019, pp. 62--69.

\bibitem{CarvalhoCMS:20}
R.~Carvalho, A.~Cunha, N.~Macedo, and A.~Santos, ``Verification of system-wide
  safety properties of {ROS} applications,'' in \emph{{IEEE/RSJ} Int. Conf. on
  Intelligent Robots and Systems ({IROS})}.\hskip 1em plus 0.5em minus
  0.4em\relax {IEEE}, 2020.

\bibitem{BrunelCCM:18}
J.~Brunel, D.~Chemouil, A.~Cunha, and N.~Macedo, ``The electrum analyzer: Model
  checking relational first-order temporal specifications,'' in
  \emph{{ACM/IEEE} Int. Conf. on Automated Software Engineering ({ASE})}.\hskip
  1em plus 0.5em minus 0.4em\relax {ACM}, 2018, pp. 884--887.

\bibitem{PurandareDED:12}
R.~Purandare, J.~Darsie, S.~G. Elbaum, and M.~B. Dwyer, ``Extracting
  conditional component dependence for distributed robotic systems,'' in
  \emph{{IEEE/RSJ} Int. Conf. on Intelligent Robots and Systems ({IROS})},
  2012, pp. 1533--1540.

\bibitem{SharmaED:17}
N.~Sharma, S.~G. Elbaum, and C.~Detweiler, ``Rate impact analysis in robotic
  systems,'' in \emph{{IEEE} Int. Conf. on Robotics and Automation ({ICRA})},
  2017, pp. 2089--2096.

\bibitem{WitteT:18}
T.~Witte and M.~Tichy, ``Checking consistency of robot software architectures
  in {ROS},'' in \emph{{IEEE/ACM} Int. Workshop on Robotics Software
  Engineering ({RoSE})}, 2018, pp. 1--8.

\bibitem{HuangEZMLSR:14}
J.~Huang, C.~Erdogan, Y.~Zhang, B.~M. Moore, Q.~Luo, A.~Sundaresan, and
  G.~Rosu, ``{ROSRV:} runtime verification for robots,'' in \emph{Int. Conf. on
  Runtime Verification ({RV})}, 2014, pp. 247--254.

\bibitem{FerrandoC0AFM:20}
A.~Ferrando, R.~C. Cardoso, M.~Fisher, D.~Ancona, L.~Franceschini, and
  V.~Mascardi, ``{ROSMonitoring}: {A} runtime verification framework for
  {ROS},'' in \emph{Towards Autonomous Robotic Systems ({TAROS})}, ser. Lecture
  Notes in Computer Science, vol. 12228.\hskip 1em plus 0.5em minus 0.4em\relax
  Springer, 2020, pp. 387--399.

\bibitem{AdamLJS:16}
M.~S. Adam, M.~Larsen, K.~Jensen, and U.~P. Schultz, ``Rule-based dynamic
  safety monitoring for mobile robots,'' \emph{Journal of Software Engineering
  for Robotics}, vol.~7, no.~1, pp. 120--141, 2016.

\bibitem{NetoASSV:19}
T.~Neto, R.~Arrais, A.~Sousa, A.~Santos, and G.~Veiga, ``Applying software
  static analysis to {ROS:} the case study of the {FASTEN} european project,''
  in \emph{Iberian Robotics Conf. - Advances in Robotics ({ROBOT})}, ser.
  Advances in Intelligent Systems and Computing, vol. 1092.\hskip 1em plus
  0.5em minus 0.4em\relax Springer, 2019, pp. 632--644.

\end{thebibliography}

\end{document}